\documentstyle[prl,aps,epsf]{revtex}
\large
\begin{document}
\twocolumn

\title{Single-Pulse Preparation of the Uniform Superpositional 
State\\ used in Quantum Algorithms}

\author{{G.P. Berman$^1$, F. Borgonovi$^{2,3}$, F.M. Izrailev$^4$, 
and V.I. Tsifrinovich$^5$}}

\address{$^1$T-13 and CNLS,  
Los Alamos National Laboratory, Los Alamos, New Mexico 87545}
\address{$^{2}$Dipartimento di Matematica e Fisica, Universit\`a Cattolica,
via Musei 41, 25121 Brescia, Italy}
\address{$^{3}$I.N.F.N., Sezione di Pavia and I.N.F.M., 
Gruppo Collegato  di Brescia}
\address{$^{4}$Instituto de Fisica, Universidad Autonoma de Puebla, Apdo.
Postal J-48, Puebla 72570, Mexico}
\address{$^5$IDS Department, Polytechnic University,
Six Metrotech Center, Brooklyn NY 11201}

\maketitle

\begin{abstract} 
We examine a single-pulse preparation of the uniform 
superpositional wave function, which includes all basis states, 
in a spin quantum computer. 
The effective energy spectrum and the errors generated by this pulse 
are studied in detail. 
We show that, in spite of the finite width of the energy 
spectrum bands, amplitude and phase errors can be made reasonably small. 
\renewcommand{\baselinestretch}{1.656} 
\end{abstract}

1.~Both the Shor and the Grover quantum algorithms begin with the  preparation 
of a uniform superposition of the basis states. 
In Shor's algorithm, it is a superposition in the $x$-register 
for the modular exponentiation: $a^x (\bmod N)$. 
For Grover's algorithm, it is a superposition of all 
possible entries of the unsorted data.
In the language of computer science, 
the transformation of the ground state of the $L$-qubit register,
$$
|0_{L-1}0_{L-2}...0_10_0\rangle,\eqno(1)
$$
into the uniform superposition of all possible basic states,
$$
\Psi_{unif}={{1}\over{2^{L/2}}}\sum|n_{L-1}n_{L-2}...n_1n_0\rangle,
~(n_k=0,1),\eqno(2)
$$
is provided by the Hadamard transformation \cite{1}. 
In physical systems, this transformation can be implemented, 
for example, using two different methods:
1) by a selected $\pi/2$-pulse excitation of each qubit, and 
2) by non-selective excitation of all qubits using a single $\pi/2$-pulse. 
The first method could be used, for example, 
for a chain of spins connected by the Ising interaction. 
Unfortunately, in this case non-resonant effects disturb 
the uniform superposition. 
The reason is that a $\pi/2$-pulse acts not only on the chosen resonant 
spin but also on all other spins \cite{2}. 
Besides, this method requires the application of $L$ pulses. 
The second method requires only a single $\pi/2$-pulse. 
This method also cannot provide a perfect uniform superposition (2). 
The second method is used currently in developed statistical 
ensemble quantum computation \cite{3,4,5}.

In this paper, we analyze the second method. First we discuss the Hamiltonian 
of the system, then the effective energy spectrum in the rotating 
reference frame and, finally, the error generated by a  $\pi/2$-pulse. 
We present the results of numerical simulation of a chain which includes 
10 spins. 
We show that, in spite of the finite width of the 
energy spectrum bands, the amplitude and phase errors can be made 
acceptably small. 

2.~Consider a chain of spin 1/2 nuclei described by the operators $I_k$. 
Assume that these spins have slightly different Larmor 
frequencies, $\omega_k$ and are connected by Ising interactions. 
In a liquid NMR quantum computation, one utilizes a statistical 
ensemble of such chains \cite{3,4,5}. 
To prepare a uniform superposition (2), 
a $\pi/2$-pulse must be polarized along the $(-y)$-axis of the rotating 
reference frame (if initially the nuclear spins point in the 
positive $z$-direction). 
The Hamiltonian of the system in the rotating frame is \cite{2},
$$
{\cal H}=\sum_{k=0}^{L-1}[-(\omega_k-\omega)I^z_k+\Omega I^y_k]-2J
\sum_{k=0}^{L-2}I^z_kI^z_{k+1},\eqno(3)
$$
where $\hbar=1$, $\Omega$ is the amplitude of the 
pulse in the frequency units (the Rabi frequency), 
$\omega$ is the frequency of the pulse, 
and $J$ is the constant of the Ising interaction. 
During the action of a $\pi$/2-pulse, the second term, 
$\Omega I^y_k$, in the Hamiltonian (3) is the main one, 
the two other terms are supposed to provide small corrections to the spin 
dynamics. To choose the values of parameters, 
we assume that the following inequalities are 
satisfied for our spin quantum computer,
$$
J\ll|\omega_{k+1}-\omega_k|,~\Omega\ll\omega_k.\eqno(4)
$$
Assuming $J/2\pi\sim 0.1$kHz, 
$(\omega_{k+1}-\omega_k)/2\pi\sim 1$kHz, 
$\omega_k/2\pi\sim 100$MHz, 
we shall consider values of $\Omega/2\pi$ up to $10$MHz. 

3.~Next, we shall discuss the effective energy spectrum 
described by the Hamiltonian (3). 
To understand the behavior of the energy spectrum, we first analyze 
a system containing two spins only. 
When the inhomogeneity and the Ising interaction 
are absent, the energy spectrum consists of three lines: 
$E_0=-\Omega$, $E_1=0$, and $E_2=\Omega$. 
The first level corresponds to the state, $|00\rangle_{-y}$, 
where the index ``-y'' indicates that both spins point in the 
(-y)-direction ($(-y)$-representation). 
The twice degenerate level, $E_1$, corresponds 
to the states, $|01\rangle_{-y}$ and $|10\rangle_{-y}$, 
and the energy level $E_2$, refers to the state $|11\rangle_{-y}$. 
First, we consider the effect of inhomogeneity 
when the Ising interaction is absent. 
Assume, for example, that $\omega=\omega_0$ 
and $\omega_1=\omega_0+\Delta\omega$. 
Then, the effective field for spin ``1'' in the frequency units is,
$
\Omega_{eff}=\sqrt{\Omega^2+\Delta\omega^2}.
$
The energy levels are given by the expressions,
$$
E_0=-(\Omega+\Omega_{eff})/2,~E_1=(\Omega-\Omega_{eff})/2,\eqno(6)
$$
$$
E^\prime_1=(\Omega_{eff}-\Omega)/2,~E_2=(\Omega+\Omega_{eff})/2.
$$
The main effect is the 
splitting of the central line into two lines with the energy separation,
$$
\Delta E=E^\prime-E\approx(\Delta\omega)^2/2\Omega.
$$
Important fact is that this splitting 
decreases as $1/\Omega$ as $\Omega$ increases.

Let us consider the error generated by a non-resonant spin. 
If the initial state of this spin is $|0\rangle_z$, then with 
linear accuracy, $\Delta \omega/\Omega$, 
the wave function of the spin, $\Psi(t)$, can be written as,
$$
\Psi(t)=[\cos(\Omega t/2)+
i(\Delta\omega/\Omega)\sin(\Omega t/2)]|0\rangle_z+\sin(\Omega t/2)|1\rangle_z.
$$
After a $\pi/2$-pulse ($\Omega t=\pi/2$), we have:
$$
\Psi(\pi/2\Omega)\approx 
{{1}\over{\sqrt{2}}}\Bigg(e^{i\Delta\omega/\Omega}|0\rangle_z+|1\rangle_z\Bigg).
$$
Thus, in first order in $\Delta\omega/\Omega$, 
the inhomogeneity generates only a phase error which decreases as $1/\Omega$. 

Now, we consider the effect of the Ising interaction 
when the inhomogeneity is absent. 
At first sight, the Ising interaction produces an 
additional ``effective field'' in the $z$-direction, 
and its influence on the energy spectrum 
must also decrease as $\Omega$ increases. 
\begin{figure}
\epsfxsize 6cm
\epsfbox{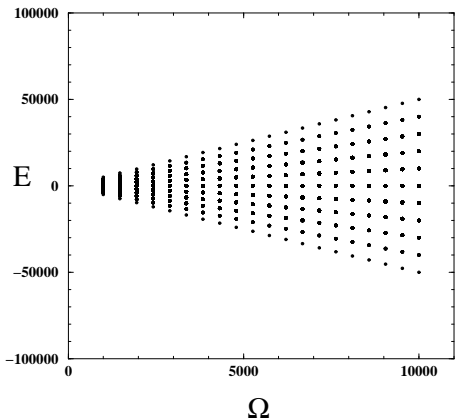}
\narrowtext
\caption{
The energy spectrum as a function of $
\Omega$.
}
\label{fig1}
\end{figure}

However, the ``effective field'' approach is not correct in this case. 
It is easy the check that the ground state is $|00\rangle_{-y}$ 
with a small admixture of the state $|11\rangle_{-y}$. 
It has the energy, $E_0=-\sqrt{\Omega^2+(J/2)^2}$.
The central energy splits into two levels $\pm J/2$, which correspond to 
symmetric and antisymmetric superpositions of the states $|01\rangle_{-y}$ 
and $|10\rangle_{-y}$, and $E_2=-E_0$. 
When the Rabi frequency, $\Omega$, increases, 
the influence of the Ising interaction on the energy levels, 
$E_0$ and $E_2$ decreases. 
But the splitting between the central 
energy levels does not change as $\Omega$ increases: 
$\Delta E_1=J$. One might expect that this splitting will 
generate an error which does not decrease as $\Omega$ increases.
Fortunately, it does not happen. 
If both spins point initially in the positive $z$-direction, the wave function, $\Psi(t)$, can be written as,
$$
\Psi(t)\approx (1/2)[e^{i\Omega t}|00\rangle_{-y}-
e^{-i\Omega t}|11\rangle_{-y}
$$
$$ 
+ie^{iJ t/2}(|01\rangle_{-y}+|10\rangle_{-y})],
$$
where we neglected the 
terms $\sim(J/\Omega)^2$. At the end of the $\pi/2$-pulse, we have,
$$
\Psi(\pi/2\Omega)
\approx{{i}\over{\sqrt{2}}}
\Bigg[|00\rangle_{-y}+|11\rangle_{-y}+
e^{i\pi J/4\Omega}(|01\rangle_{-y}+|10\rangle_{-y})\Bigg].
$$
\begin{figure}
\epsfxsize 6cm
\epsfbox{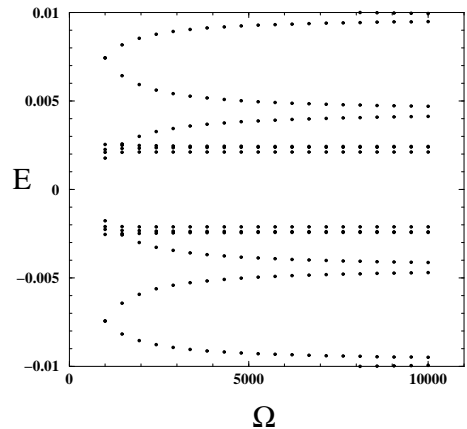}
\narrowtext
\caption{The structure of a narrow strip of the central band shown in
Fig.\ref{fig1}.}
\label{fig1c}
\end{figure}

For $J=0$, we have a uniform superposition of the basis states 
(in both $(-y)$- and $z$-representations). 
For $J\not=0$, to first order in $J/\Omega$, 
the Ising interaction (similar to the inhomogeneity) 
generates only a phase error which decreases as $1/\Omega$. 
The non-vanishing bandwidth, $\Delta E_1=J$, does not cause 
a non-vanishing error. 
This happens because the phase error is proportional to the duration 
of a $\pi/2$-pulse which is proportional to $1/\Omega$. 

3.~Next, we  present the results of numerical simulations 
with $L=10$ qubits. 
These simulations require operations on a digital computer 
in the Hilbert space with dimension $D=2^{10}=1024$. 
\begin{figure}
\epsfxsize 6cm
\epsfbox{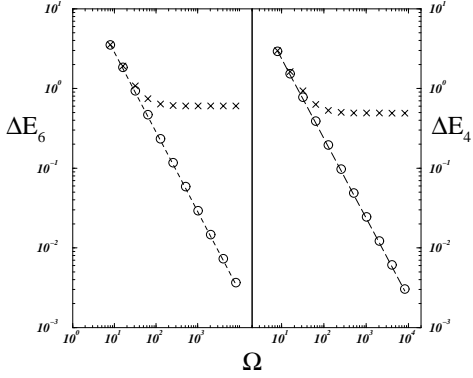}
\narrowtext
\caption{Widths $\Delta E_6$,(left) $ \Delta E_4$ (right)
 of the $6$th (central) and the 4th band vs $\Omega$.
Open circles are for $J=0$, crosses for $J=0.1$.
Dashed lines are  the best fit to $A/\Omega$ (circles) where 
$A=28.466$ (left) and $A=23.673$(right).
}
\label{bandnew}
\end{figure}

If the Ising interaction and the inhomogeneity of the frequencies, 
$\omega_k$, are both absent, the energy spectrum of the Hamiltonian (3)  
consists of 11 equidistant levels separated by  gaps with value $\Omega$. 
Both the interaction and the inhomogeneity cause splitting of all inner levels. 
This leads to formation of energy bands. 

Fig. \ref{fig1} shows the energy spectrum 
as a function of $\Omega$, for the following values of parameters,
$$
\omega_{k+1}-\omega_k=1,~(k=0,...,9),~J=0.1,~\omega_k-\omega=k-4.5.
$$

\begin{figure}
\epsfxsize 6cm
\epsfbox{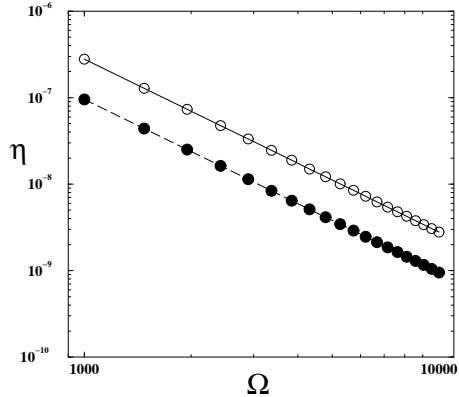}
\narrowtext
\caption{Dependence on $\Omega$ of the maximal 
(open circles) and average errors (full circles
) for the amplitude modulus, $|A_n|$, on $\Omega$.
The full line is the best fit, $0.2787/\Omega^2$, while the dashed is 
$0.0953/\Omega^2$.
Other parameters are the same as in
Fig.\ref{fig1}.}
\label{ampli}
\end{figure}

The last equation means that the frequency $\omega$ of a $\pi/2$-pulse is equal
to the average Larmor frequency, 
$\langle\omega_k\rangle=(1/10)\sum_{k=0}^{9}\omega_k$. 
For each $\Omega$, there are 1024 energy levels.  
At the scale shown in Fig. \ref{fig1}, each band is represented by a point. 

\begin{figure}
\epsfxsize 6cm
\epsfbox{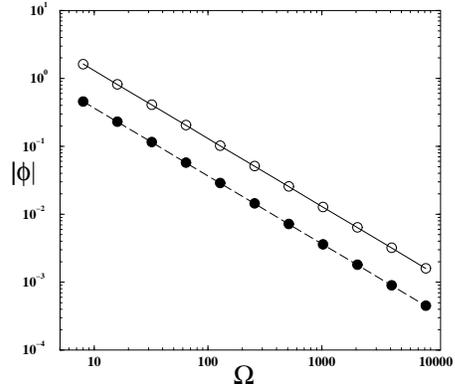}
\narrowtext
\caption{Dependence of the maximal (open circles) and average errors (full circles
) for the phase errors $|\phi_n|$, on $\Omega$.
Full line is the best fit $13.0216/\Omega$, while dashed is
$3.6606/\Omega$.
Other data are the same as in
Fig.\ref{fig1}.}
\label{pha}
\end{figure}

In fact, each band has a complicated structure. As an example, Fig. \ref{fig1c}
 shows the structure of a narrow strip of the central band 
as a function of $\Omega$.
Fig. \ref{bandnew} (left) shows the dependence of the width of the 
central band, $\Delta E_6$, on $\Omega$ (in the logarithmic scale). 
When the interaction between spins is absent (circles in 
Fig. \ref{bandnew} ), $\Delta E_6$ decreases approximately 
as $1/\Omega$. A finite interaction (crosses in Fig. \ref{bandnew}) 
changes this picture. 
After the width of the band caused by the inhomogeneity 
decreases to the value of approximately $6J$ 
(at $\Omega\approx 50$), its value does not decrease. 
This dependence of $\Delta E_6$ on $\Omega$ 
is qualitatively similar to the results discussed above for two spins. 
Fig. \ref{bandnew} (right)  shows similar dependence for the the width 
of the 4th band, $\Delta E_4$. 

\begin{figure}
\epsfxsize 6cm
\epsfbox{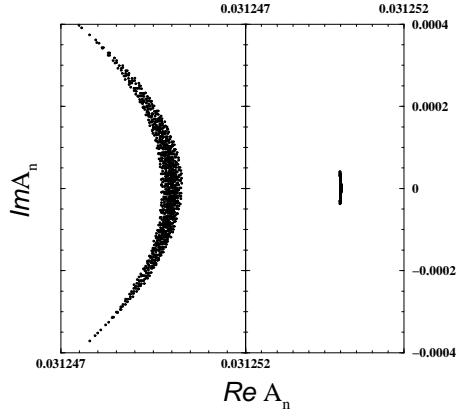}
\narrowtext
\caption{
The distribution of the complex amplitudes, 
$A_n$, in the complex plane for $\Omega=10^3$ 
(left) and $\Omega=10^4$ (right).
Other parameters are the same as in
Fig.\ref{fig1}.}
\label{camp}
\end{figure}

Now we consider the errors generated in the process of preparation of a 
uniform superposition of the basis states, $\Psi_{unif}$. 

 In the rotating frame, the spin dynamics 
can be described as a superposition of stationary 
solutions with constant coefficients, which can be found 
from the initial conditions. 
In the absence of both the interaction between spins and the 
inhomogeneity, a $\pi/2$-pulse applied to the ground state, 
$|0_{L-1}...0_0\rangle_z$, generates a uniform superposition of 
all $2^L$ basic states with the amplitudes $1/\sqrt{2^L}$. 

\begin{figure}
\epsfxsize 6cm
\epsfbox{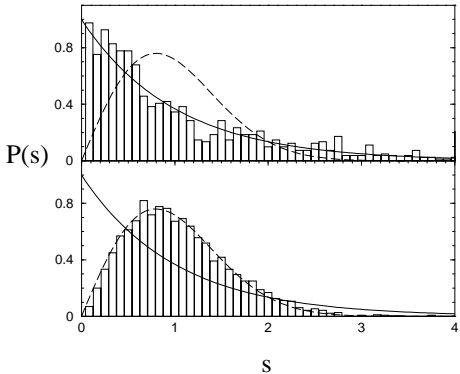}
\narrowtext
\caption{Level spacing distribution for a system of $L=12$ spins; 
$\Omega=100$; $\omega=\omega_0$, $\omega_k=\omega_0+k$,
 $J=20$ (upper); $J=100$ (lower).  
The continuous line shows the 
Poisson distribution, while the dashed curve shows the GOE distribution. 
}
\label{pofs}
\end{figure}

To describe the error in the complex amplitude, 
$A_n=|A_n|\exp(i\Phi_n)$, 
of the state $|n\rangle_z$ 
$(0\le n\le 2^{L-1})$, we use two quantities: 
1) $\eta=|2^{-L/2}-|A_n||$ which describes the error of the amplitude modulus,
 $|A_n|$, 
and 2) the phase modulus, 
$|\Phi_n|$, which describes the phase error (as $\Phi_n=0$ in the ideal case). 
Fig. \ref{ampli} shows the dependences of the 
maximal error, $\eta_{max}$ and the average error, $\eta_{ave}$, 
on the Rabi frequency, $\Omega$. 
One can see that both quantities decrease approximately as $1/\Omega^2$. 
Fig. \ref{pha} shows similar dependences of the phase error on $\Omega$. 
Both, $|\Phi|_{max}$ and $|\Phi|_{ave}$ decrease approximately as $1/\Omega$. 
This corresponds to the results derived above for two spins. 
Note that for parameters we chose (which correspond to $J\ll\Delta \omega$), 
the main contribution to the error depends on the inhomogeneity. The 
plots presented in Figs \ref{ampli} and \ref{pha} do not change 
significantly when $J=0$. 
Fig. \ref{camp} shows the distribution of the complex 
amplitudes, $A_n$, in the complex plane for $\Omega=10^3$ and 
$\Omega=10^4$. 
One can see that the distribution of $A_n$ has a 
form of the arc whose length and width decrease as $\Omega$ increases.

4.~In this section, we  note that the quantum 
Hamiltonian (3) belongs to the so-called quantum non-integrable systems. 
This means that, treated classically, these systems exhibit chaotic 
behavior for some range of parameters and initial conditions. 
(See, for example, \cite{6} and references therein.) 
The quantum properties of classically chaotic systems can be investigated 
by a detailed analysis of their eigenvalues and eigenfunctions\cite{frep}.
This analysis will be the subject of future investigation.
Here, we concentrate on the statistics of neighboring energy levels spacing,
$P(s)$. Indeed it has been conjectured that $P(s)$ for a chaotic system 
depends only on some general symmetries\cite{bohi}. 
In our case (for Hermitian matrices), 
 we can  assume the quantum chaos regime as 
described  by a Wigner-Dyson distribution.
Fig. \ref{pofs} shows the transition  of  
$P(s)$, from the Poisson distribution 
to the Wigner-Dyson distribution of the Gaussian orthogonal ensemble (GOE) 
of Hermitian matrices. 

As one can see from Fig. \ref{pofs}, the transition to chaos 
appears for relatively large values of $J$ which are far beyond 
the range of quantum computation.
Also, the appearance of the Poisson distribution, which should be 
a fingerprint of integrability, appears at relatively high values of
$J$, out of quantum computation regime.
Level statistics in the quantum computation regime will be the 
subject of future investigations.

5.~In conclusion, we investigated the errors generated by a 
single-pulse implementation of the Hadamard transformation for 
a chain of spins connected by the Ising interaction. 
In the rotating reference frame, the interaction between spins 
and the inhomogeneity of the Larmor frequencies split the 
energy levels into the bands. 
The characteristic width of the band caused by the inhomogeneity 
decreases as $1/\Omega$ as the Rabi frequency, $\Omega$, increases. 
The contribution to the band widths due to the Ising 
interaction between spins remains constant. 
In spite of this fact, errors generated in the process of 
preparation of the uniform wave function decrease monotonically 
as $\Omega$ increases. 
When the Rabi frequency, $\Omega$, increases, the errors of the 
amplitude modulus, $|A_n|$, decrease as $1/\Omega^2$. 
The phase errors decrease only as $1/\Omega$. 
For reasonable values of $\Omega$, the errors of the amplitude's 
modulus, $|A_n|$, become negligible. 
The phase error can be reduced to the order of $10^{-5}$rad. 
Thus, both the errors caused by the inhomogeneity and the interaction 
between spins can be made reasonably small, for a single-pulse 
generation of the superpositional wave function used 
in the main quantum algorithms. 
\section*{Acknowledgments}
The work  of GPB and VIT was supported by the Department of 
Energy (DOE) under contract
 W-7405-ENG-36, by the National Security Agency (NSA) and Advanced Research 
and Development Activity (ARDA). FB acknowledges financial support from INFN and INFM.
\end{document}